\documentclass[11pt,twoside]{article}  % Leave intact
\usepackage{adassconf}

\begin{document}

\paperID{O3.4}
\title{Closing the loop: Linking Datasets to Publications and Back}

\author{Alberto Accomazzi, G\"unther Eichhorn, Arnold Rots}
\affil{Harvard-Smithsonian Center for Astrophysics, 60 Garden St.,
Cambridge, MA, 02138, USA}

\contact{Alberto Accomazzi}
\email{aaccomazzi@cfa.harvard.edu}
\paindex{Accomazzi, A.}
\aindex{Eichhorn, G.}
\aindex{Rots, A.}
\authormark{Accomazzi, Eichhorn \& Rots}

\begin{abstract}
With the mainstream adoption of references to datasets in astronomical
manuscripts, researchers today are able to provide direct links from
their papers to the original data that were used in their study.
Following a process similar to the verification of references
in manuscripts, publishers have been working with the NASA
Astrophysics Data System (ADS) to validate and maintain links to these
datasets.

Similarly, many astronomical data centers have been tracking 
publications based on the observations that they archive, and have
been working with the ADS to maintain links between their datasets
and the bibliographic records in question. In addition to providing
a valuable service to ADS users, maintaining these correlations allows
the data centers to evaluate the scientific impact of their missions.

Until recently, these two activities have evolved in parallel on
independent tracks, with ADS playing a central role in bridging the
connection between publishers and data centers. However, the ADS is
now implementing the capability for all parties involved to find out
which data links have been published with which manuscripts, and vice
versa. This will allow data centers to periodically harvest the ADS
to find out if there are new papers which reference datasets available
in their archives. In this paper we summarize the state of the
dataset linking project and describe the new harvesting interface.
\end{abstract}

\section{Introduction}

Scientists responsible for a mission or a data archive know 
very well how important it is to keep track of who is using their data 
and how it is being used.
Especially in this day and age of intense competition for research 
funds, it has become essential for large projects to maintain a 
list of scientific works that have been published based on the 
project's data, and to evaluate the project impact in terms 
of bibliometric measures.
Partly due to this motivation, many data centers have begun 
maintaining publication lists in electronic form since the 
first days of the web.  
This data has been shared with the ADS so that proper linking 
can occur between the ADS and the data centers.  

In 2004 the AAS journals have introduced the option for 
authors to reference the datasets used in their studies in
the body of their papers, with the purpose of strengthening 
the linking between publications and the supporting data.
The ADS has once again been involved in facilitating this effort,
by implementing a scheme that allows the automated verification
of dataset identifiers during the copy-editing process, and the
automated linking of the identifiers from the electronic
manuscripts.

More recently, the ADS and the Chandra Data Archive have been
prototyping a service that allows a data center
to harvest the list of links published in the literature
which correspond to the datasets it archives.  
This allows a data center to supplement and verify the
list of links it maintains between its data product and the
published literature and provides a new paradigm for 
sharing linking information among collaborating institutions.

\section{Links between the Literature and Datasets}

Links between datasets maintained by the 
data centers and the ADS bibliographic records to
which they are related have existed since 1995.
These links have been maintained by librarians or
data archivists as lists of bibcodes and URLs,
and they have been shared with the ADS to allow
the generation of cross-links.
The creation and maintenance of such links has
become an increasingly important aspect of the data
archives operations (see, e.g. Rots et al. 2004).

In addition to providing access to 
publications related to a certain dataset, having a list
of bibliographic records for a mission or project
has also allowed the ADS to define bibliographic 
groups for some of the major missions.
For instance, one can query the ADS for all papers 
published on Chandra data, and then 
get the total number of citations for them.
As the data centers have become the authoritative repositories of the 
links between their datasets and the 
corresponding bibliographic records, the
ADS has acted as an intermediary 
between the journal articles and the data.

In an attempt to improve the quality and usefulness of
astronomical manuscripts, the Astrophysics Data Center Executive 
Committee (ADEC) in 2004 
endorsed the proposal of creating links in the AAS journals to
datasets maintained by the main astrophysics
data archives (Eichhorn et al. 2004).
The proposal calls for the definition of 
Dataset Identifiers according to the following syntax:
{\em ivo://ADS/FacilityId\#PrivateId}.
These identifiers are designed to address a broad range of
data granularity, from individual observations to collections of datasets, 
and may correspond to either static collection of data products
or to data generated on the fly by a service
(see, e.g., Alexov \& Good, 2006).
They are defined as a type of IVO Identifiers (Plante et al. 2006),
and as such they are permanent, unique, 
and resolvable.

The implementation of the linkage from published papers to datasets
requires the collaboration of both the data centers and the publishers, 
and is mediated by the ADS (Accomazzi \& Eichhorn 2004).
Data centers assign identifiers to the datasets they serve,
uniquely identifying them.  These identifiers need to be prominently
displayed in the metadata related to the dataset, so that
the scientists analyzing it will know how 
to reference it during publication.
Publishers who participate in this effort provide suitable
means for authors to include a reference to the data during the 
creation of the manuscript (typically via the use of LaTeX macros),
and to verify the validity of the datasets through a central 
verification service provided by the ADS.
The central verifier forwards the query to the appropriate 
data center(s) in order to determine if a certain dataset is valid
and what its final linking URL should be.
The results of this query are cached in a local database and are
used later to resolve the dataset identifier when a user 
requests it.

\section{Harvesting of Dataset links}

The current linking schemes, described above, provide the basic
functionality desirable from a user's perspective: they allow
access to the publications relevant to a particular dataset
(when accessing the dataset); and they allow access to the 
datasets described in a scientific paper (when accessing 
an online article).  However, this level of linking relies
on two independent tracks, each headed by a different 
group of people: the data centers and the publishers.
While these procedures accomplish their original goals, 
the current exchange of metadata among the parties involved 
has so far not facilitated the curation of links from the data
centers' point of view, since they have not received information
about what new data links are being published in the
current literature.

When a new article containing references to datasets is published,
the relevant metadata is made available by the publisher to the ADS
in the form of a list of 
dataset identifiers which have been successfully verified
and resolved to a particular online data product.
The ADS has now created a harvesting interface that data centers
can use in order to discover which of the datasets in their
archive are being referenced by new journal articles.
The current interface supports simple HTTP
REST-based queries returning straight XML, 
but we expect to extend it to support
additional access methods as requested by the VO community.
The interface supports both the complete and 
incremental retrieval of records related to a particular observing
facility or maintained by a particular data center.

The harvesting procedure described above allows data centers
to maintain an updated list of the datasets referenced by the 
current literature.  
We don't expect this to be an exhaustive list
for a variety of reasons: 
some publications don't (yet) participate
in this effort, so no linking metadata will be available from them; 
authors can't always be trusted to properly cite all of the
datasets they have analyzed in their study;
and the verification procedures used to determine the validity of a link
may have failed due to transient problems.  
For these and other reasons, we consider the data centers to be the
authoritative source of links between publications and data products.
Therefore the ADS will continue to 
rely on the correlation tables provided by the 
data centers to create links between bibliographic 
records and datasets.  To this end, all data centers maintaining
these correlations should provide a harvesting interface that
will be used by ADS to populate the corresponding 
links from its bibliographic records
back to the data centers.

As an added benefit, this harvesting interface can serve as a
prototype for a more general 
service allowing the exchange of relational tables
between any two sets of identifiers (in this case bibcodes and 
dataset identifiers).  A similar service could be used by two
data centers to share related links between them, e.g. 
to maintain lists of coordinated observations.

\section{Conclusions}

We have reviewed the existing procedures that allow the
exchange of metadata used in creating links between 
the different entities involved in maintaining bibliographies
and astronomical data products.  
We have proposed an improvement of the current linking procedures
based on the use of harvesting services that can facilitate
the curation and exchange of 
links between bibliographies and datasets.
Even with the adoption of our recommendations, much work
still needs to be done by all parties involved in order to 
further the creation and maintenance of these links.  
In particular, these are the areas where we think most
our effort should be concentrated on:

\begin{itemize}
\item Data centers should make users aware of the identifiers 
that have been assigned to the datasets they download and analyze,
so they can be referenced in the literature.  
Most of the NASA data centers have already provided this 
information to their users, but we hope for greater participation
from the international community at large.

\item Publishers should encourage 
the scientists to declare in their papers which
datasets have been used in their study and reference them
using the appropriate editing tools.  The AAS journals have
led the way in this effort, and we hope that the remaining
astronomical journal publishers will join in the coming years.

\item Data centers should harvest the ADS to find out what
papers have published links to datasets under their curation.
The data centers should also provide ADS with a complete
list of links between their data products and the published
literature.
\end{itemize}

We hope that broader participation in this important project 
can be accomplished by raising the awareness of these efforts
within the astronomical community.  More information about
this project is available online at \\
\htmladdURL{http://vo.ads.harvard.edu}

\acknowledgments

The ADS is funded by NASA Grant NNG06GG68G.
The Chandra X-Ray Center is funded  by NASA Grant NAS 8-03060.


\begin{references}

\reference Accomazzi, A.\ \& Eichhorn, G.\ 2004, \adassxiii, 181 

\reference Alexov, A.\ \& Good, J.\ 2006, \adassxvi, \paperref{P3.01}

\reference Plante, R., Linde,~T., Williams, R., \& Noddle,~ K. 2006, \\
\htmladdURL{http://www.ivoa.net/Documents/latest/IDs.html}

\reference Rots, A.~H., Winkelman, S.~L., Paltani, S., Blecksmith, S.~E., \& Bright, J.~D.\ 2004, \adassxiii, 605 

\reference Eichhorn, G., \& Astrophysics Datacenter Executive Committee 
(ADEC) 2004, Bulletin of the American Astronomical  Society, 36, 805 

\end{references}
\end{document}